\documentstyle[12pt,aasms4]{article} 

\def\n253{NGC 253 }
\def\deg{$^\circ $ }
\def\kms{\mbox{$\rm kms^{-1}$}}
\def\zx1{$\rm x_{1}$}
\def\x2{$\rm x_{2}$}
\def\ha{H$\alpha$~}
\def\sec{$^{\prime\prime}$}
\def\arcsec{\prime\prime}
\def\anantha{Anantharamaiah \& Goss (1996) }
\def\pv{P-V }

\begin{document}

\title{The Central Velocity Field in NGC 253 : Possible Indication of a Bar } 

\author{Mousumi Das\altaffilmark{1, 2},  K.R. Anantharamaiah\altaffilmark{2, 3}
               \& M.S. Yun\altaffilmark{3}}
\affil{\altaffilmark{1}Indian Institute of Astrophysics, Koramangala, 
                      Bangalore 560 034, India}
\affil{\altaffilmark{2}Raman Research Institute, C.V Raman Avenue, 
                    Sadashivanagar, Bangalore 560 080, India}
\affil{\altaffilmark{3}National Radio Astronomy Observatory, Socorro, 
                  NM 87801, USA}

\centerline{e-mail: mousumi@astro.umd.edu, anantha@rri.ernet.in, myun@daisy.astro.umass.edu}

\begin{abstract}

We have investigated whether motion of gas in a bar-like potential can
account for the peculiar but systematic velocity field observed in the
nuclear region of the starburst galaxy NGC 253. This unusual velocity
field with gradients along both major and minor axes was revealed in a
high resolution ($1.8^{\arcsec}\times 1.0^{\arcsec}$) H92$\alpha$ 
recombination line observation by
Anantharamiah and Goss (1996).  A simple logarithmic potential is used
to model the bar. Assuming that the bulk of the gas flows along closed
and non-intersecting \zx1 (bar) and \x2 (anti-bar) orbits of the bar
potential, we have computed the expected velocity field and 
position-velocity diagrams and compared them with the observations.

A comparison of the integrated CO intensity maps with the spatial 
distribution of the \zx1 and \x2 orbits in the model indicates that the 
nuclear molecular gas in NGC 253 lies mainly on the \x2 orbits. We   
also find that the velocity field observed in the central 100 pc region
in the H92$\alpha$  recombination line is well accounted for by the bar model
if most of the ionized gas resides in the inner \x2 orbits.  However, the
model is unable to explain the velocity field on a
larger scale of $\sim 500$ pc observed using the OVRO interferometer
with a resolution of $5^{\arcsec}\times 3^{\arcsec}$. The direction of
the observed CO velocity field appears twisted compared to the
model.  We suggest that this perturbation in the velocity field may be due
to an accretion event that could have occurred $10^7$ years ago. 

\end{abstract}

\section{INTRODUCTION}

NGC 253 is a nearby, barred Sc galaxy with a nuclear starburst region
(${\rm L}\sim 3 \times 10^{10}{\rm L}_{\odot}$; Telesco \& Harper
1980).  The galaxy has been extensively studied at different
wavelengths.  The proximity of the galaxy (d$\sim$ 3.4 Mpc) and the
ongoing nuclear starburst makes it an ideal candidate to study
enhanced star formation at nearby distances. The galaxy is nearly edge
on ($i=78^{o}$) with a considerable amount of dust in the center which
makes it bright in the infrared wave bands. Observations of the
H$\alpha$ emission line indicate possible nuclear outflows and stellar winds in
the central region, arising from enhanced star formation (Ulrich 1978,
Schulz \& Wegner 1992). Radio continuum observations reveal numerous
compact sources within the inner 200 pc which are either supernova
remnants or HII regions (Ulvestad \& Antonucci, 1997).  At
near-infrared wavelengths (NIR), a prominent bar can be observed in
the center (Scoville et al. 1985) whose position angle is tilted by
about 18\deg with respect to the major axis of the galaxy
. Interferometric CO observations with a resolution of $\sim 5''$ have
shown the presence of a molecular bar of dimensions
$\rm30^{\arcsec}\times10^{\arcsec}$, whose orientation is similar to
the NIR bar (Canzian, Mundy \& Scoville, 1988). HCN and CS emissions,
which trace dense gas, also reveal a similar bar ( Paglione et al 1995
and Peng et al. 1996). Recent work on the bar of NGC 253 indicate that
there is at least one Inner Lindblad Resonance (ILR) point for the
galaxy (Arnaboldi et al. 1995).  Various types of data on NGC 253
indicate that the nuclear starburst is mainly located within a
circumnuclear ring in the bar (e.g. Engelbracht et al. 1998).

The kinematics of gas in the nuclear region of \n253 has been known
to be anamolous from the early \ha observations by Demoulin
\& Burbidge (1970). Radial velocity measurements at various position
angles near the nucleus show that, in addition to solid body rotation,
large non-circular motions exist within the central region (Ulrich 1978,
Shulz \& Wigner 1992, Munoz-Tunon, Vilchez \& Castaneda 1993). However,
due to problems of obscuration at optical wavelengths, results from
\ha and [NII] observations are difficult to interpret. Longer wavelength
observations such as that of CO (Canzian, Mundy \& Scoville 1988),
Br$\gamma$ (Puxley and Brand 1995) and H$_2$ (Prada et al. 1996) indicate
steeper velocity gradients along the major axis and only solid body
rotation in the central region. Most of the optical and IR
observations, aimed at studying the kinematics of the nuclear region of
\n253 (e.g. Arnaboldi et al. 1995, Prada et al. 1996, Prada, Gutierrez
\& McKeith 1998, Engelbracht et al. 1998),  have relied on measuring the
velocity gradients along chosen position angles passing though the
nucleus (e.g. major axis and minor axis). However interferometric
radio observations at millimeter and centimeter wavelength (e.g.
Canzian et al. 1988, Paglione et al. 1995, Anantharamaiah \& Goss 1996,
Peng et al. 1996) measure two-dimensional velocity
fields which provide a more complete  picture of the kinematics in the
nuclear region. These two-dimensional measurements reveal a complex
but systematic velocity pattern in the central region. In addition to
solid body rotation, the observed velocity fields indicate motions
which may be due to a bar-like potential in the center (Peng et al
1996) or a kinematic sub-system which may be caused by a past merger
event (Anantharamaiah \& Goss 1996).

The highest resolution measurement of the two-dimensional velocity
field available to date is that of Anantharamiah \& Goss (1996) who
observed the H92$\alpha$ recombination line with a beam of
$\rm1.8^{\arcsec}\times1.0^{\arcsec}$ and a velocity resolution of 54
\kms. The line emission is detected in the central
$\rm9^{\arcsec}\times4^{\arcsec}$ (approximately 150 pc$\times$60 pc)
region of \n253 and oriented roughly along the major axis of the
galaxy.  The observed velocity field of the H92$\alpha$ line emission
shown in Figure 1a, has an elongated S-shaped pattern with
iso-velocity contours running almost parallel to the major-axis of the
galaxy. For pure solid body rotation, the iso-velocity contours are
expected to run parallel to the minor-axis. \anantha showed that this
velocity field could be fitted to a set of three orthogonally rotating
nested rings of ionized gas. Such an interpretation is however only
empirical and has no clear physical basis. The central velocity field
of \n253 has been observed over a larger scale
($\rm\sim30^{\arcsec}\times 10^{\arcsec}$) but with coarser angular
resolution ($\rm\sim3^{\arcsec}-6^{\arcsec}$) in molecular lines; CO
(Canzian et al. 1988), HCN (Paglione et al. 1995) and CS (Peng et al
1996).  As mentioned earlier the molecular line emission is oriented
along the NIR bar which is tilted by 18\deg with respect to the major
axis whereas the H92$\alpha$ line emission is mainly along the major
axis.  The velocity field of the molecular gas is also distinctly
different from that of the ionized gas.  Peng et al. (1996) have
suggested that the morphology and kinematics of the dense molecular
gas can be explained by gas moving in a bar potential.
  
In this paper, we have attempted to model the inner velocity field of
NGC 253 to determine whether the velocity pattern observed by \anantha
can also be explained by motion of gas in a bar potential. We also
present in this paper, new observations of the velocity field in the
CO line with an angular resolution of $5.6''\times 2.6''$.  Since gas
is a collisional system (unlike stars which can be collisionless), it
tends to settle on closed, non-intersecting orbits in the bar
potential.  Using a bar model, we have detemined a set of closed
orbits from which a velocity field is derived and compared with
observation in the H92$\alpha$ and the CO lines.  While the bar model
succesfully accounts for the velocity field observed in the
H92$\alpha$ line in the inner ($9''\times 4''$) region, it does not
account of the observed CO velocities on a larger scale ($40''\times
15''$).  The only earlier attempt to model the nuclear velocity field
of NGC~253 using a bar model is that of Peng et al (1996) who
attempted to explain the velocity field observed on the larger scale
in the CS line. Although, Peng et al (1996) have reported success of
their model, by comparing observed position-velocity (PV) diagrams
with PV plots of closed x1 and x2 orbits, we show in this paper that a
single bar model cannot explain both the H92$\alpha$ and the CO (or
the CS) velocity fields.

\section{New Observations of the CO Velocity field}

In Figure 1b, we show a new measurement of the velocity field of
$^{12}$CO obtained using the Owens Valley Radio Interferometer. The
angular resolution is $\rm5.6^{\arcsec}\times2.6^{\arcsec}$.  These
observations were made as a part of a detailed study of molecular gas
in the nuclear region of NGC 253, which will be published elsewhere
(Yun et al., in preparation). A comparision of Figs 1a and 1b shows
that there are several differences in the velocity fields observed in
the H92$\alpha$ and the CO lines.  While the CO emission occurs over a
much larger region ($\sim 40''\times 15''$), the H92$\alpha$ line is
confined to the inner most region ($\sim 9''\times 4''$). The position
angles of the two emission regions are different: while the
H92$\alpha$ line emission is along the major axis of the galaxy (PA =
52\deg), the CO emission is oriented along the axis of the NIR bar (PA
= 70\deg).  The iso-velocity contours in Figures 1a and 1b run along
entirely different position angles.  Some distortion in the CO
velocity field is seen at the position where the H92$\alpha$ emission
is observed. It is possible that, at higher angular resolution, the CO
velocity field may resemble the H92$\alpha$ velocity field in the
central region.  The iso-velocity contours in Figure 1b also do not
run parallel to the minor axis of the galaxy indicating that the
kinematics of the CO gas is not dominated by standard galactic
rotation. The velocity field observed by Peng et al. (1996) in the CS
line is similar, but less systematic compared to the CO velocity field
in Figure 1b.

\section{Motion of Gas in the Bar Potential}

Gas clouds dissipate energy through collisions. Hence such clouds will
tend to move along closed orbits in a plane. In a bar potential, there
are two types of closed orbits in the plane of the galaxy; the \zx1
(bar) orbits which are extended along the major axis of the bar and
the \x2 (anti-bar) orbits which are oriented perpendicular to the major
axis of the bar (Contopoulos \& Mertzanides, 1977, Athanassoula, 1988,
Binney \& Tremaine, 1987). The \x2 orbits exist if  there is an Inner
Lindblad Resonance (ILR) ring in the galaxy,  which occur at radii 
where the relation $\Omega(R)-\Omega_{b}=\kappa(R)/2$ is satisfied. Here
$\Omega(R)$ is the angular speed of the particle, $\Omega_b$ is the
angular speed of the bar and $\kappa$ is the epicyclic frequency.
Physically, this means that a particle rotating in the bar potential
with angular speed $\Omega(R)$, will encounter successive crests of the
bar potential at a frequency equal to half its epicyclic frequency.
For radii close to the ILR
radius, a particle leads the bar and for radii near the Outer Lindblad
Resonance (OLR), the particle lags behind the bar.  At radii
where $\Omega(R)-\Omega_{b}=0$, a particle is stationary in the
rotating potential. More than one ILR can exist; in fact from the
optical rotation curve of NGC 253, Arnaboldi et al. (1995) conclude
that there are two ILRs in the bar of NGC 253, one at a radius of
25\sec and the other close to the center. 

At the intersection of the \zx1 and \x2 orbits, gas clouds may
collide, lose angular momentum and sink into the \x2 orbits.
Simulations of gas evolution in bars have shown that an evolved bar
has relatively more gas on the inner orbits than on the outer orbits
(Friedli \& Benz, 1993). Within the bar of NGC 253, there is a $\sim
20''$ diameter nuclear ring of dense star forming gas (Arnaboldi et al
1995) which may imply that there is a significant amount of gas on \x2
orbits within the bar. We therefore use the \zx1 and \x2 orbits to
model the gas flow in the center of NGC 253. This method of explaining
gas kinematics in a bar potential was first introduced by Binney et
al. (1991) for the distribution of gas in the Galactic Center. Since
then, this method has been applied to many barred galaxies
(e.g. Achtermann \& Lacy, 1995). This approach was used by Peng at
al. (1996) to explain the distribution of dense gas in the bar of NGC
253. Peng et al. (1996) plotted the position velocity (P-V) diagram
for closed orbits in the bar model and compared it with the observed
P-V diagram for the CS line emission. Based on this comparison, they
interpreted the regions of CS emission as the apocenters in the \zx1
and \x2 orbits of the bar potential.

An alternative approach to determine the velocity field is the technique
of hydrodynamic simulation of gas flow in a bar potential (e.g. Piner, 
Stone and Teuben, 1995, Athanassoula, 1992). In hydro simulations it 
is possible to follow the evolution of the gas in the bar and to trace
the formation of shocks in the gas. However, the closed orbits method
which is used in this paper, gives an explicit picture of the 
distribution of the gas. The distribution of the orbits shows the 
regions where gas can settle (the orbits themselves) and the regions 
where gas will pile up in shocks (i.e. at the ends of the bar and the 
crossing points of the x1 and x2 orbits). In this method, the velocity 
field can also be explicitly obtained and compared with the observations. 
 
Although we also adopt a bar model and the closed-orbits method,
which is similar to that of Peng et al.  (1996), our approach differs
in at least two ways.  First, in constructing the bar model, Peng et
al. (1996) used a logarithmic bar potential and scaled all length
scales with the disk rotation velocity $\rm v_b$. When comparing the
closed orbits with the observed distribution of CS intensity and the
P-V plot, Peng et al. (1996) scaled the orbits with $\rm v_{b}=75$
\kms. However, the velocity in the flat portion of the rotation curve
is $\approx$ 200 \kms (Arnaboldi et al.  1995). The value $\rm v_b=75$
\kms\- leads to a bar structure in which the \zx1 orbits extend out to
$\sim$20\sec\- resulting in a bar size of $\leq$45\sec.  However, IR and
optical observations estimate a bar size of $\sim$150\sec\-
(e.g. Scoville et al. 1985).  Our bar model incorporates parameters
from the rotation curve and hence leads to a more realistic bar
size. In the bar model presented in the following sections, we have
used $\rm v_{b}$=200 \kms. This velocity gives a bar size of
$\approx$200\sec, which is closer to the observed value than that used
by Peng et al. (1996). The large size of the bar may indicate that not
all of the outer \zx1 orbits support gas and hence star formation;
most of the gas might have been funnelled into the inner orbits.
 
The second difference between the work presented here and that of Peng
et al (1996) is that, in addition to comparing the model and observed
P-V diagrams, we construct a model two-dimensional velocity field and
compare it with the observed velocity fields shown in Figs 1a and
1b. We show that an explicit comparison of the predicted and model
velocity fields is indeed essential to determine whether the gas is
actually moving on regular \zx1 and \x2 orbits. Because of the
inherent degeneracy in projection, a wide range of models can fit the
\pv plot, whereas the added dimension in the 2-D velocity field helps
differentiating the models.

\subsection{Bar Potential and Parameters}

For simplicity we have also adopted a logarithmic bar potential to
model the velocity field. The potential is given  by,
\begin{equation}
\Phi(x,y) = \frac{1}{2}(v_b)^{2}{\rm ln}(x^2 + \frac{y^{2}}{q^{2}} + R_c^{2})
\end{equation}
\noindent (Binney \& Tremaine, 1987), where $v_b$ is the velocity in
the flat portion of the rotation curve of the galaxy, q is the
non-axisymmetry parameter and $\rm R_c$ is the core radius. If the bar is
rotating with an angular velocity $\Omega_b$, then the equation of
motion of a particle moving in the rotating frame of the bar is given
by
\begin{equation}
\ddot{\bf r} = -\nabla\Phi - 2({\bf{\Omega_b}}\times{\bf v}) - 
{\bf{\Omega_b}}\times({\bf{\Omega_b}}\times {\bf r}) ,
\end{equation}
\noindent where {\bf r} is the position vector of a particle in the bar, 
{\bf v} is the velocity of a particle in the rotating frame of the
bar and $\ddot{\bf r}$ is the acceleration.

The closed orbits in the bar potential were determined for different
values of q and $\rm R_c$. Some of the outer \zx1 orbits which are
elongated along the length of the bar, are looped i.e. self
intersecting at the ends.  Since gas clouds cannot survive in these
orbits due to collisions at the intersection points, these orbits were
excluded while determining the velocity field.  The free parameters
were chosen such that the number of looped orbits is minimum.  The
parameters defining the bar potential are $v_b$, $\Omega_b$, q and
$\rm R_c$. Of these four parameters, $v_b$ and $\Omega_b$ can be
determined from the rotation curve. These parameters were taken from
the optical rotation curve observed by Arnaboldi et al. (1995). The
velocity in the flat portion of their rotation curve is $\rm v_b \sim
200 ~ \kms$ and the angular rotation velocity of the bar is $\Omega_b
=$ 48 \kms kpc$^{\rm -1}$. The adapted parameters of the galaxy are
given in Table 1.

Closed orbits were determined for values of the the non-axisymmetry
parameter q in the range 0.7 to 0.9. For lower values of q, most of
the closed orbits are looped.  Even for q=0.7, most of the outer \zx1
orbits are looped at the ends. For q=0.9, the bar structure is not as
pronounced as that for the q=0.7 \& 0.8 models but the non-axisymmetry
is clearly seen. For both q = 0.8 and 0.9, we obtain a large range of
non-looped \zx1 and \x2 orbits.  The models presented in this paper
are for q=0.8 \& 0.9. For both values of the parameter, there is a
region of overlapping \zx1 and \x2 orbits. At these points the clouds
collide, lose angular momentum and sink into the center. This process
is thought to be important for transporting gas into the center of the
galaxy (Binney et al. 1991, Friedli \& Benz 1993).

The other free parameter in the bar model, core radius $\rm R_c$ , was
varied from 0.05 pc to a few hundred parsecs. For $\rm R_c$ less than
a few parsecs, the velocity field resembles that of a simple flat
rotation curve, i.e. it appears like the `spider diagram' observed in
the disks of galaxies. For $\rm R_c~>~$200 pc, distinct \zx1 and \x2
orbits are not obtained.  The range $\rm 5~ pc~<<~R_c~<~$200 pc gives
a bar structure and a velocity field which is different from that expected from
normal galactic rotation. We examined the velocity fields for values
of core radii $R_c = $50 pc, 100 pc and 150 pc and  found them to be
similar. In the following sections, results are presented for $\rm
R_c$=100 pc. 

\subsection{Bar and Anti-Bar Orbits}

Figures 2a \& 2b show the closed orbits in the bar potential in the plane of
the galaxy as they appear when the galaxy is viewed face on. For these
orbits, the non-axisymmetry parameter q=0.8 and  0.9 respectively 
and the core radius $\rm R_c = 100$ pc.
The closed orbits were also determined for the same
value of q and $\rm R_c$ = 50 pc and 150 pc. The overall appearance of
the bar is the same for all the three cases. The extent of the \zx1 orbits
is approximately 3.2 kpc ( $3'.2$) in all the cases and their
orientation is also similar.  Since gas will settle along non-looped
\zx1 and \x2 orbits in the bar potential, the radial extent of the
bar in this model is 3.2 kpc. This radial extent is
comparable to the size of the NIR bar observed by Scoville et
al. (1985).  The inner \x2 orbits extend out to about 400 pc ($\sim
25''$) along the semi-major axis.  The \x2 orbits corresponding to
different core radii do differ from one another. For core radii 100 pc
and 150 pc, the \x2 orbits within 20 pc are aligned perpendicular to
the other \x2 orbits i.e. along the \zx1 orbits. The result is a bar
within a bar structure.  However, if the orbits are convolved with a
two dimensional Gaussian representing the telescope beam, these
innermost perpendicular orbits, which are lying within the \x2 orbits,
are no longer distinct.  Thus in the final convolved velocity fields,
there are no significant differences between models with 
core radii lying in the range  $\rm 5~ pc~<<~R_c~<~$200 pc.

In order to compare the model velocity field with the observations,
the closed orbits were projected on to the plane of the sky.
Projections were made using the parameters given in Table 1.  If the
bar orbits are projected on the plane of the sky, the size and
orientation of the \zx1 and \x2 orbits change significantly as shown
in Figures 3a \& 3b. The \zx1 orbits are aligned along the bar (PA
$=70^{o}$).  The \x2 orbits do not appear to be perpendicular to the
\zx1 orbits.  Instead, these orbits are at a position angle of
$45^{o}$ on the plane of the sky which is $7^{o}$ with respect to the
major axis of the galaxy.  This inner ring appears like a narrow ridge
when projected on the plane of the sky even though it has a radius of
$\sim 350$ pc.  In the plane of the sky, the inner ring is also nearly
aligned with the major axis of the galaxy rather than the axis of the
bar. This alignment is due to the projection of the \x2 orbits on the
sky and has been discussed by others (e.g. B$\ddot{\rm o}$ker, Krabbe
\& Storey, 1998). This inner ring can be identified with the ring of
star forming regions and supernova remnants which have been observed
both in the optical and radio bands (Arnaboldi et al. 1995, Baan et
al. 1997). A large number of compact radio sources have been observed
in the inner 200 pc of NGC 253 which are dentified as supernova
remnants and HII regions (Ulvestad \& Antonucci 1997). A high
concentration of dense molecular gas is also observed in the inner
ring within the bar (e.g. Peng et al. 1996).

A comparison of the extent of the projected bar orbits in Fig 3, with
the observed extents of CO, CS and HCN emissions, shows that the
molecular gas must lie mainly on the \x2 orbits. The molecuar gas lies
within a radius of 30\sec\- from the center, within which, only one (the
inner most) \zx1 orbit is present (Fig 3).  Within the region of
molecular emission, the computed orbits are mainly of the \x2-type for
both q=0.8 and 0.9. All the molecuar line observations thus indicate
that gas has been channelled in to the inner most orbits inside the
bar and the resulting high concentration of gas may have given rise to
a burst of nuclear star formation. We note that in the bar model
considered by Peng et al (1996), the molecuar gas is present in the
outer \zx1 orbits as well. This difference is because, Peng et al
(1996) have implicitly assumed a smaller size for the bar.

\section{Comparison with Observations}

\subsection{The Velocity Field}

To obtain the velocity field expected in the bar model and compare it to the
observations, the radial velocity at each point along
the closed orbits was determined by tranforming the velocity
components in the rotating frame of the bar to the inertial frame
using
\begin{equation}
{\bf v_{in}}={\bf v(x,y)} + {\bf\Omega_{b}}\times{\bf r}.
\end{equation}
\noindent The radial velocity is given by
\begin{equation}
{\bf v_{rad}}=[(v_{x}-\Omega_{b}y)sin\phi + (v_{y}+\Omega_{b}x)cos\phi]sin(i).
\end{equation}
\noindent Hence, for every point {\bf r(x,y)} along the closed 
orbit, there is a corresponding radial velocity ${\bf v_{rad}}$.
Using the above equation, it is possible to plot isovelocity contours for
the radial velocities of the closed bar orbits projected onto the
plane of the sky. To make a proper comparison with the observations, 
the model velocity field is convolved with an appropriate 
two dimensional Gaussian representing the telescope beam. The comparison 
between model and observed velocity fields has the limitation that while the 
observed velocity field is "weighted" by the intensity of line emission,
the model velocity field is based only on the distribution of closed orbits.
In other words, the model implicitly assumes that the gas is uniformly
distributed over all the closed orbits.

Figure 4a shows the model velocity field within the inner 8\sec derived
from the closed orbits shown in Figure 2a (q=0.8). Figure 5a is the 
corresponding velocity field for the orbits shown in Figure 2b (q=0.9). The 
convolving Gaussian
function has a size $\rm1.8^{\arcsec}\times 1^{\arcsec}$ which is similar
to the resolution used by \anantha in their H92$\alpha$ observations.
A comparison of  Fig 1a with Fig 3 reveals a remarkable
correspondence between the observed and model velocity fields in the
central region for both q values. The model velocity field predicts isovelocity 
contours which are nearly parallel to the major axis as observed. The S-shaped
pattern in the observed field is also present to some extent in the
model velocity field. It thus appears that most of the ionized gas
observed in the H92$\alpha$ line by \anantha can be associated with
the \x2 orbits of the bar potential.

The velocity gradients along the major and minor axes of the galaxy in
the inner region of the model can be quantitatively compared with the
values obtained by \anantha.  For q=0.8 and $\rm R_c=50$ pc to 150 pc,
the bar model predicts a velocity gradient along the minor axis of
$\sim$25 to 30 $\rm kms^{-1}arcsec^{-1}$ and the gradient along the
major axis is $\sim $10 to 15 $\rm kms^{-1}arcsec^{-1}$.  \anantha
measured a gradient of $\sim$18 $\rm kms^{-1}arcsec^{-1}$ along the
minor axis and $\sim$11 $\rm kms^{-1}arcsec^{-1}$ along the major axis
of the galaxy. The estimates from the model are thus comparable to the
observed values.

Figure 4b shows the model velocity field for the parameter q=0.8, on a
larger scale of $\rm54^{\arcsec}\times$30\sec and Figure 5b is the
corresponding velocity field for q=0.9. These figures should be
compared with the CO velocity field shown in Figure 1b. The angular
resolution is 5\sec$\rm \times$3\sec. The region covered in the
figures includes mainly all of the \x2 orbits. The model velocity
field appears somewhat different for the two q values. The q=0.9
contours have a closer resemblance to the observed velocity field
(Figure 1b) than the q=0.8 model. The velocity field lines for q=0.9
are more spread out than the q=0.8 model and beyond 5\sec have a
downward slope which matches fairly well with the observed field. To
see if the velocity field in Figure 1b could be reproduced by
including only a subset of the orbits, we first constructed the velocity
field with just the \x2 orbits and then with the \x2  and a few 
of the inner \zx1 orbits. In both cases, we were not able to construct
a velocity field similar to that observed in CO and other molecular
lines.

We thus come to the conclusion that the model of the bar potential can
only explain the velocity field close to the nucleus and not on the
scale of $45^{\arcsec}$ shown in Figure 1b.  We conjecture that there
is some perturbation within the bar which causes a change in the
direction of the bar at progressively larger radii. Such a change in
the position angle of the bar with radial distance has in fact been
observed by Baan et al (1997) in the formaldehyde absorption
line. They conclude that there must be a warped gas disk in the
nuclear region which results in the changing direction of the bar with
radius.  However, the gross appearance of the bar does not seem to be
affected by the perturbed velocity field.  In Secion 4 we consider a
possible scenario to explain the difference between the observed and
model CO velocity fields. Our conclusion, based on a comparison of the
velocity field is thus different from that of Peng et al (1996) who
compared the PV diagarms and concluded that the bar model can account
for the velocities observed on a larger scale in molecular lines. In 
the next section we carry out such a comparison of the PV diagrams.

\subsection{Position-Velocity Diagram}

The closed orbits in a bar potential can also be used to construct a 
\pv diagram and has been done for NGC 253 and a few other galaxies in 
the literature (e.g. Peng at al. 1996, Achtermann \& Lacy 1995). But
the resultant \pv plot has two major differences with the observed \pv
diagram. First, in the model \pv plot, V is just the rotational
velocity of the gas in the bar potential (V$_{\rm 0}$); the velocity
spread $\rm\Delta v$ is not incorporated in the model. If the gas is
cold, like molecular hydrogen, then $\rm\Delta v$ may be relatively
small and this difference may not be important. But if the gas is hot
and turbulent, then $\rm\Delta v$ may be fairly large. Secondly, the
model \pv plot does not include information about the intensity (I) of
the line emission from the gas. The second problem can be partially overcome by
assuming that the intensity is proportional to the gas distribution in
the bar. Thus if the gas is uniformly distributed over the closed
orbits, then the intensity is proportional to the number of orbits
crossing through that region. We gridded the \pv plane so that each
grid point had a set of coordinates (P,V,I).  The size of the grid
unit was chosen to be much smaller than the beam of the telescope.
For example, in the inner 9\sec of NGC 253, we used a grid unit size
of 0.3\sec$\times$8 \kms. The ``beam''  used to observe
that region is 1.6\sec$\times$54 \kms (\anantha). This grid or 2-d
matrix was then convolved with a 2-d Gaussian matrix representing the
beam of the telescope.

Figure 6a is the  \pv diagram of the nuclear ionized gas in
NGC 253 observed in the H92$\alpha$ radio recombination line by \anantha.
There is a large velocity spread in the center of about 400 \kms and
two secondary peaks are seen on either side of the central peak.  A
velocity gradient ($\sim$ 10 \kms ) is clearly seen.  Figure 6b is
the beam-convolved model \pv diagram of the inner 8\sec constructed
from the \zx1 and \x2 orbits, as discussed above.  Considering the
simplicity of our model, there is a reasonable agreement between the 
two plots. Although the intensity contours in the model are in
arbitrary units, there is a
overall similarity which includes the presence of a main peak and 
two secondary peaks, although the latter are not as pronounced as in the
observed P-V plot. Furthermore, the velocity gradient is in the right
sense. The velocity spread
near the central peak and the magnitude of the velocity gradient are,
however, less than the observed values in Figure 6a.  Some of the
differences may be caused by the fact that the observed \pv plot is
constructed from the H92$\alpha$ radio recombination line which is
emitted from hot, possibly turbulent, ionized gas for which the
velocity spread could be quite large. This velocity spread 
is not included in the model.

Figure 7a shows a P-V diagram along the major axis observed on a much
larger scale ($\sim$ 45\sec) in CO, using the OVRO synthesis array
(Sec 2). There are two strong peaks on either side of the center and a
gradient in velocity from NE to SW. Figure 7b shows the beam-convolved
model \pv diagram over a similar region for q=0.8. The \pv plots for
q=0.8 \& 0.9 are very similar and so we have shown only the q=0.8
figure in the paper. There is a reasonable similarity between the
model and observed \pv plots. The extent and gradient of the velocity
field shown in Figure 7b agrees fairly well with that in Figure 7a. On
the basis of this plot, we could conclude that the model explains the
observed kinematics of molecular gas.  Such a conclusion was in fact
arrived at by Peng et al. (1996) who compared their observed \pv
diagram (in the CS line) with the \zx1 and \x2 orbits projected on to
the \pv plane. There is also a similarity between the model \pv plots
in Fig 7b and that observed in the CS line by Peng et al. (1996).
However, as shown in the previous section, there is no similarity
between the observed CO velocity field (Figure 1b) and the model
velocity field (Figures 4b \& 5b).  We, therefore, conclude that the
velocity field observed on larger scales in molecular gas (Canzian et
al. 1988, Peng et al. 1996 and Figure 1b ) is not explained by the bar
potential, although the observed and model \pv diagrams have many
similarities.  Thus an agreement between the observed and expected \pv
diagram is a necessary but not sufficient condition to ascertain
whether a particular model accounts for the observed velocity field.
It is necessary to also compare the 2-d velocity field predicted by
the model with the observations.

A careful comparison of the larger scale ($\sim 500$ pc) observed and
model velocity fields (Figures 1b and 7b) shows that it may be possible
to interpret the observed field as a twisted version of the model
field. It is possible that the velocity field can indeed be due to
the bar potential, but an additional perturbation has altered the
velocity structure. In the next section we conjecture that this
perturbation was caused by an accretion event that occurred in the
recent past.

\section{Does the Velocity Field Reveal Evidence of a Merger ?}

NGC 253 is considered to be a strong, nuclear starburst galaxy with
 star formation concentrated within the inner \x2 ring of the bar
(Peng et al. 1996). Though a bar can produce enhanced star formation
in the center of a galaxy, the rapid fueling of gas in a bar can also be
triggered by the accretion of a small satellite galaxy (Mihos \& 
Hernquist, 1994). That such an accretion event may have occurred in NGC
253 has been suggested by \anantha and Prada, Gutierrez \& McKeith
(1998).

In the previous section, we showed that the nuclear velocity field can
be explained reasonably well by gas moving on \x2 orbits within the
bar. But as we move out to radii of $\sim$30\sec and farther, we find
that there is no agreement with the observations. This disagreement 
is possibly due to the change in the position angle of the bar which 
has been observed on scales of $\sim $1 kpc 
(Baan et al. 1997). Such a perturbation of the bar
potential could have been caused by the postulated merger event in the
recent history of the galaxy. However, since the velocity field at
large distances from the starburst region is regular and the outer HI
contours are also regular (Combes 1977, Puche 1991), the merger event
has not caused any disruption on a large scale.  Therefore, if there
was a merger in the recent past, then the accreted mass must be fairly
small compared to the mass of the galaxy.

The perturbation in the velocity field is observed at radii of
$\sim$30\sec and larger, which corresponds to gas moving on the outer
\x2 orbits in the bar. The gas in the inner $\sim$100 pc, which is
moving along inner \x2 orbits, does not seem to have been perturbed
since the observed velocity field is similar to that predicted by the
model. This behavior may be explained by the difference in the
rotation velocities for gas moving along the outer and inner \x2
orbits. The rotation time scales for gas moving in the inner ($<$100
pc) orbits in our model is $\sim $ few times $10^{6}$ years while that
for the outer \x2 orbits is $\sim $ few times $10^{7}$ years. Thus the
model suggests that the very inner gas is moving at least ten times
faster than the gas in the outer \x2 orbits.  Perturbations in the
central potential due to the accretion of a small galaxy may have been
smoothed out by the faster rotating inner gas but its effect may still
remain on the scale of the outer \x2 orbits where gas is rotating more
slowly. The accretion event may thus have occurred about $10^7$ years
ago.

We have used the above hypothesis to determine an approximate mass for
the accreted galaxy using the theory of dynamical friction. From
Binney \& Tremaine (1987), the deceleration of a body of mass M moving
through a region of mass density $\rho$, is given by,

\begin{equation}
\frac{dv_M}{dt}=\frac{4\pi(ln\Lambda)G^{2}\rho M}{{v_M}^{2}}[erf(X)
     -\frac{2Xe^{-X}}{\sqrt{\pi}}]  
\end{equation}

\noindent where X=$\frac{v}{\sqrt{2} \sigma}\sim 1$ and ln$\Lambda 
\sim $10 to 20 (see Binney \& Tremaine, 1987, pg 429). 
The density $\rho$ can be determined from the dynamical mass within
the inner $5^{\arcsec}$ ($\sim $150 pc), which is $\sim 3\times 10^{8}
M_{\odot}$ (Anantharamaiah \& Goss, 1996). The accreted mass is then given by,

\begin{equation}
M=\frac{{v_M}^{3}}{12\pi (ln\Lambda)G^{2}\rho [erf(X) 
   -\frac{2Xe^{-X}}{\sqrt{\pi}}]\Delta t}
\end{equation}

\noindent If we assume that $\Delta t$ is the time taken for the mass
M to sink through the inner 300 pc, then $\Delta t\sim 10^{7}$ years.
This time scale is similar to the time required for the perturbation
to have been smoothed out in the inner \x2 orbital region but the
perturbation may still be persisting in the outer \x2 orbital
region. We have taken $\rm v_M$ to be the rotation speed at 300 pc
which is $\sim$110 \kms (Arnaboldi et al. 1996). Using these values,
the mass of the accreted body M $\sim 10^{6} M_{\odot}$. This mass is
too small to represent a galaxy but it may represent the remains of an
accreted galaxy which had most of its outer gas and stars stripped off
during the infall.  In deep optical images of NGC 253, made using
image enhancement techniques (Malin 1981; Beck, Hutschenreiter and
Wielebinski 1982), a distortion in the disk is observed with a spur of
emission protruding to the south. It is possible that this distortion
is an imprint left by the merger event that occurred $10^7$ years ago
as we have hypothesized above. Also, Watson et al.  (1996) have
detected four star clusters in the inner parsecs of NGC 253, one of
which, has a mass of $\rm\sim 1.5\times 10^{6} M_{\odot}$ and the
others have masses of the order of $\rm 10^{4} M_{\odot}$. The more
massive cluster may be the remains of the accreted mass which has
perturbed the bar potential.

\section{SUMMARY}

We have attempted to model the observed nuclear velocity field of
ionized and molecular gas in NGC 253 using a simple logarithmic bar
potential. The parameters for the potential were derived from the
optical rotation curve. The velocity field was determined from the \zx1
and \x2 orbits in the bar potential and compared with observations of
the H92$\alpha$ (Anantharamaiah \& Goss 1996) and CO lines.  The
results are summarized below.

\noindent 1) The \x2 orbits, projected onto the plane of the sky, lie
roughly along the major axis of the galaxy. The \x2 ring appears very
narrow and ridge like. We identify the ionized gas observed in the
H92$\alpha$ line as lying along the inner \x2 orbits. Also, when we
compare the integrated CO intensity maps in the literature with the
model we find that the molecular gas is also distributed mainly over
the \x2 orbits.

\noindent 2)   The velocity field within the inner 8\sec
predicted from the bar potential is similar to that observed by
\anantha.  The model predicts velocity gradients both along the
major and minor axes. The isovelocity contours run parallel to
the major axis as observed. The model \pv diagram agrees
reasonably well with the observations.

\noindent 3) The model velocity field  on a larger scale ($\sim$45\sec) is
significantly different from the observed velocity fields in 
CO  and CS (Peng et al. 1996).  However, the
model \pv diagrams agree reasonably with the observations.
We therefore conclude that agreement in \pv diagrams is a necessary
but not sufficient condition to explain the complete velocity field.

\noindent4) The observed velocity field on a larger scale ( $\sim$45\sec)
appears to be a twisted version of the model velocity field with the
direction of the gradient changing with radius.  We suggest that this
perturbation may have been caused by the accretion of an object with
mass of $\sim 10^{6} M_{\odot}$ about $10^{7}$ years ago. This
accretion event may also have triggered the nuclear starburst observed
in the galaxy. Enhanced optical images of NGC 253 (Malin 1981, Beck et al
1982) show a distorsion in the outer regions of the galactic disk which may
be a signature left by the merger event.

\acknowledgments

We thank Niruj R. Mohan for useful suggestions regarding the
isovelocity contour plots and the \pv diagrams and W.M. Goss for a
critical reading of the manuscript.  The National Radio Astronomy
Observatory is a facility of the National Science Foundation operated
under cooperative agreement by Associated Universities, Inc.

\newpage

\centerline{\bf FIGURE CAPTIONS}

\noindent Figure 1 (a) Velocity Field of the central 9\sec  observed in the
H92$\alpha$ line using the VLA by \anantha, with a resolution of 
1.8\sec$\times$1.0\sec. RA and Dec offsets are with respect to the position
($\rm \alpha, \delta$)1950 00$^{h}$ 45$^{m}$ 5.80, -25 33 39.1. Contour
levels range from 70~\kms to 270~\kms in steps of 10~\kms. The grey scale 
ranges from 100~\kms to 400~\kms. 

\noindent (b) Intensity weighted mean velocity field derived from the
CO (1--0) data imaged using the OVRO synthesis array at a resolution
of 5\sec.6$\rm\times$ 2\sec.6 ($\rm PA=-2^\circ$).  The RA and Dec
offsets are with respect to the radio nucleus position
($\rm\alpha(1950)=00^h 45^m 05^s.79,~\delta(1950)=-25^\circ 33'
39''.08$).  Contour levels range from 100~\kms to 360~\kms in steps of
20~\kms. The grey scale ranges from 100~\kms to 400~\kms.

\noindent Figure 2a.  A face-on view of closed orbits in the bar
potential in the plane of the galaxy NGC 253.  The bar parameters are
$\rm v_b$=200 \kms, $\rm\Omega_b$=48 \kms$\rm kpc^{-1}$, q=0.8, $\rm
R_{c}$=100pc. The outer \zx1 orbits lie along the bar major axis and
the inner \x2 orbits lie along the bar minor axis.

\noindent Figure 2b. A face-on view of closed orbits in the bar
potential for the bar parameters $\rm v_b$=200 \kms, $\rm\Omega_b$=48
\kms$\rm kpc^{-1}$, q=0.9, $\rm R_{c}$=100pc.

\noindent Figure 3a. Closed orbits in Figure 2a projected onto the plane of 
the sky. The galaxy has a position angle of 52\deg and the PA of the
bar is 70\deg.  In this projection, the outer \zx1 orbits lie along the bar
and the inner \x2 orbits have a position angle of 45\deg.

\noindent Figure 3b. Closed orbits in Figure 2b projected onto the
plane of the sky. As in Figure 3a, the galaxy has a position angle of
52\deg and the PA of the bar is 70\deg. The outer \zx1 orbits lie
along the bar and the inner \x2 orbits have a position angle of
45\deg.
 
\noindent Figure 4a. Model radial velocity field in the central
8$^{\prime\prime}$, constructed from closed orbits in the bar
potential of Figure 2a.  The model is convolved with a Gaussian beam
1.8\sec$\times$1.0\sec, which is equal to the beam in Figure
1a. Contours are marked in units of $\rm kms^{-1}$. The radial
velocities are offset with respect to the central velocity of the
galaxy. The dashed rectangular box shows the region where the
H92$\alpha$ line is observed (see Figure 1a).

\noindent Figure 4b. Model radial velocity field in the central 8\sec,
constructed from closed orbits in the bar potential of Figure 2b.  The
model is convolved with a Gaussian beam 1.8\sec$\times$1.0\sec.
Contour levels are marked in $\rm kms^{-1}$ and as before the radial
velocities are offset with respect to the center of the galaxy.  The
dashed rectangular box shows the region where the H92$\alpha$ line is
observed (see Figure 1a).

\noindent Figure 5a.  Model radial velocity field in the central
$54^{"}\times 30^{"}$ constructed from the closed orbits in the bar
potential of Figure 2a. The model is convolved a with Gaussian
function of dimension 5\sec$\rm \times$3\sec which is the same as the
resolution in Figure 1b. Contours are marked in units of $\rm
kms^{-1}$.  The radial velocities are offset with respect to the
central velocity of the galaxy.  The dashed parallelogram shows the
region from which the CO line is observed (see Figure 1b).

\noindent Figure 5b.  Model radial velocity field in the central
54\sec$\rm \times$ 30\sec constructed from the closed orbits in the
bar potential of Figure 2b. The model is convolved with a Gaussian
function of dimension 5\sec$\rm \times$3\sec. Contours are marked in
units of $\rm kms^{-1}$ and the velocities are offset from the center
of the galaxy. The dashed parallelogram shows the region from which
the CO line is observed (see Figure 1b).

\noindent Figure 6a. Position velocity diagram of the ionized gas in
the inner 8\sec along the major axis of NGC 253.  The diagram is
constructed from the H92$\alpha$ recombination line data of
Anantharamaiah \& Goss (1996) with a spatial resolution of 1.8$''\times
1''$ and a velocity resolution of 54 \kms.  Contour levels are 2, 4, 6,
...,  24 mJy beam$^{-1}$.  Offsets along the Y axis are from $\rm
v_{Hel}$ = 200 \kms. Offsets along the X axis are from RA(1950) =
00$^{\rm h}$ 45$^{\rm m}$ 5$^{\rm s}$.9.
 
\noindent Figure 6b. Position velocity diagram of the gas
moving along the closed orbits in the inner 8\sec constructed from the
closed orbits in the bar potential of Figure 2a. The intensity is assumed to be
proportional to the orbit number density. The image is convolved with
a Gaussian function of size 1.5\sec$\times$56 \kms, which is similar to the
resolution in Figure 6a. The contour levels are in arbitrary 
units.

\noindent Figure 7a. Position-Velocity plot of the CO (1--0) emission
along the morphological major axis of NGC~253 ($PA=51^\circ$) observed
using the OVRO synthesis array.  The major axis offset is with respect
to the radio nucleus position ($\rm \alpha, \delta$)1950 00$^{h}$
45$^{m}$ 5.80, -25 33 39.1.  and the velocity offset at zero
corresponds to the LSR velocity of +239 km sec$^{-1}$.  The CO line
intensity is shown both in greyscale and in contours, which are linear
increments of 0.4 Jy beam$^{-1}$ ($2\sigma$). The velocity resolution is 
20.8 \kms.

\noindent Figure 7b. Model \pv diagram constructed from closed orbits
of Figure 2a for the inner 45\sec. The intensity is assumed to be
proportional to the number density of orbits.  The convolving function
has size of 6\sec$\times$20 \kms, which is similar to the beam
resolution in Figure 7a. The contour levels are in arbitrary units.

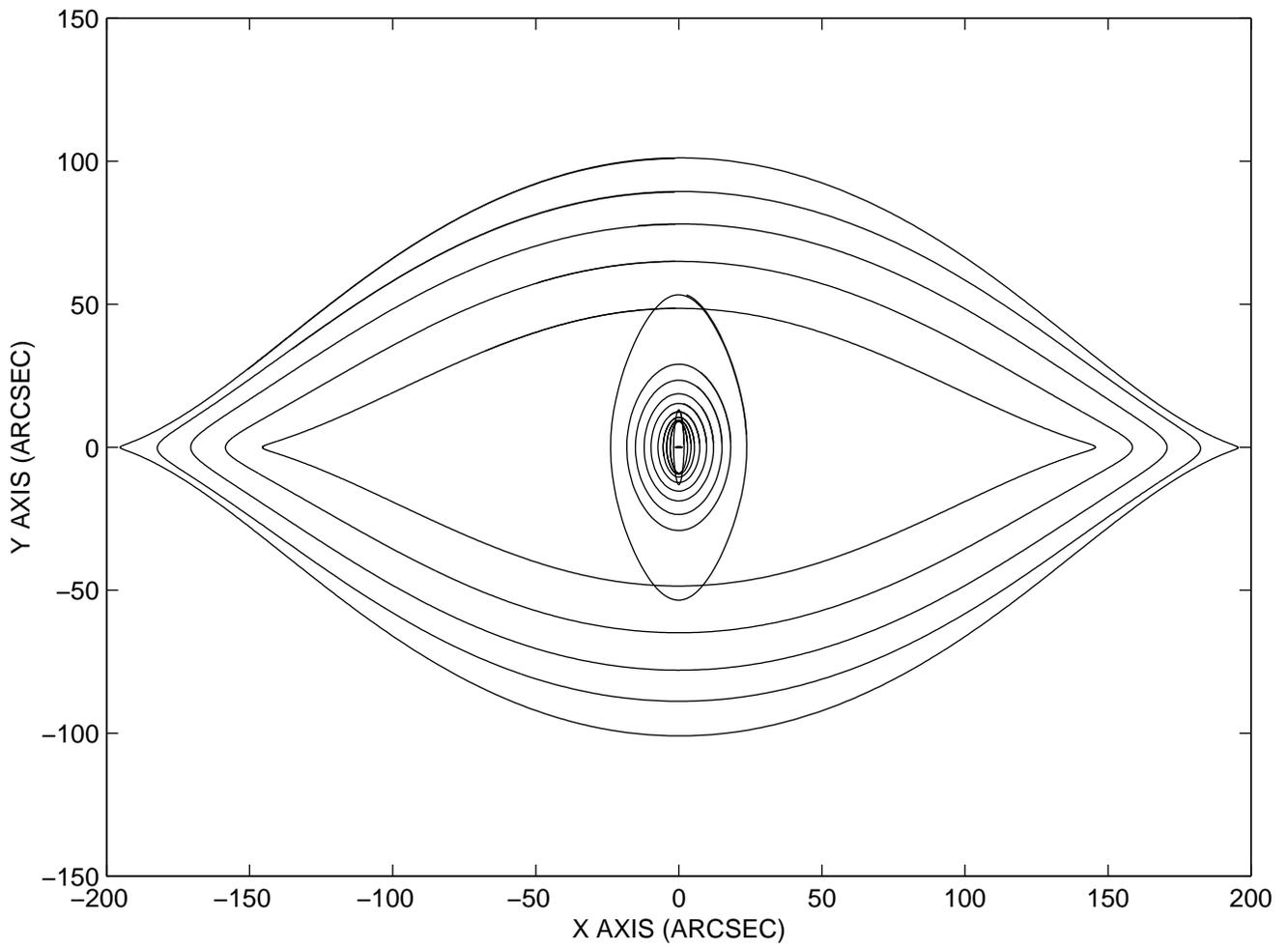
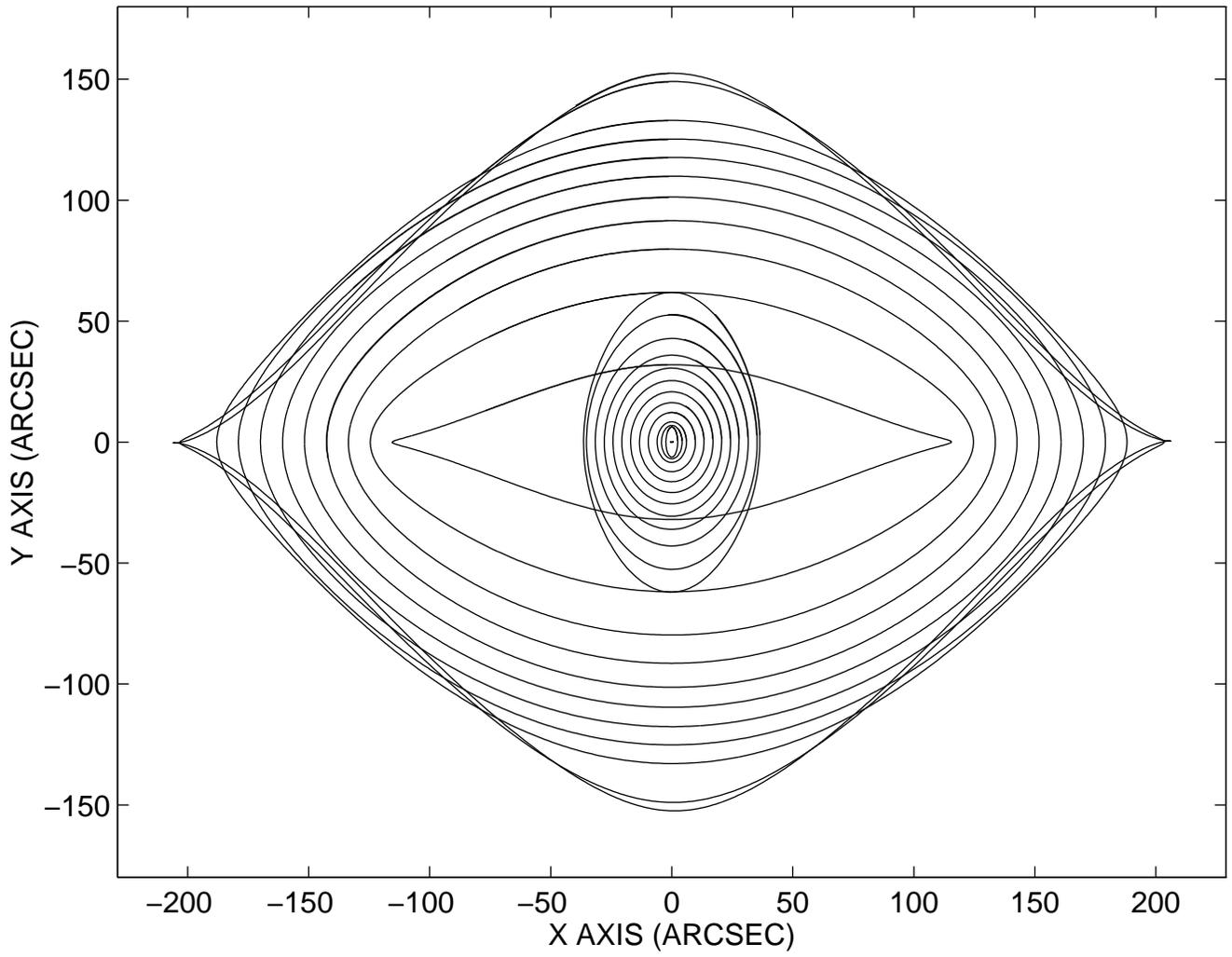
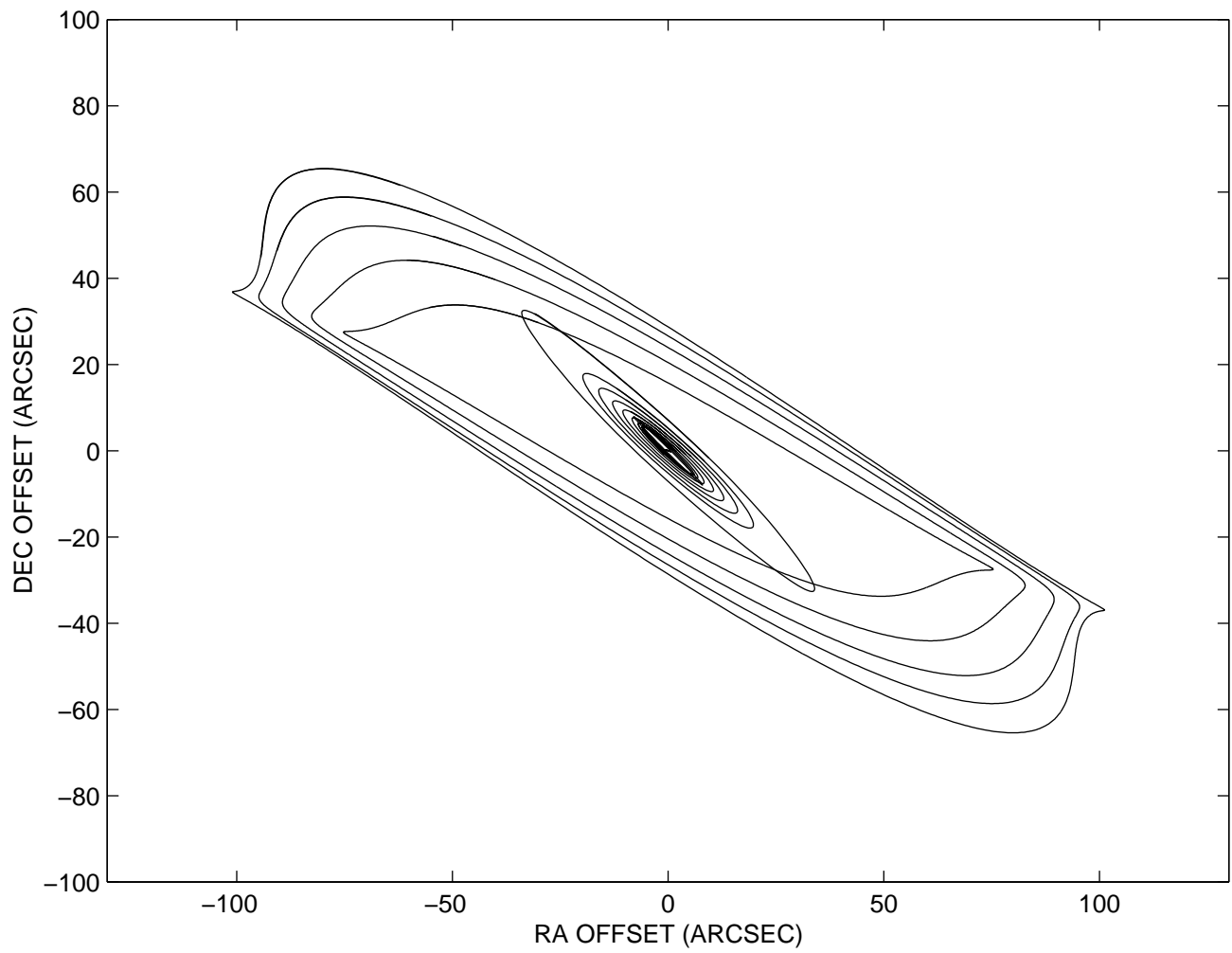
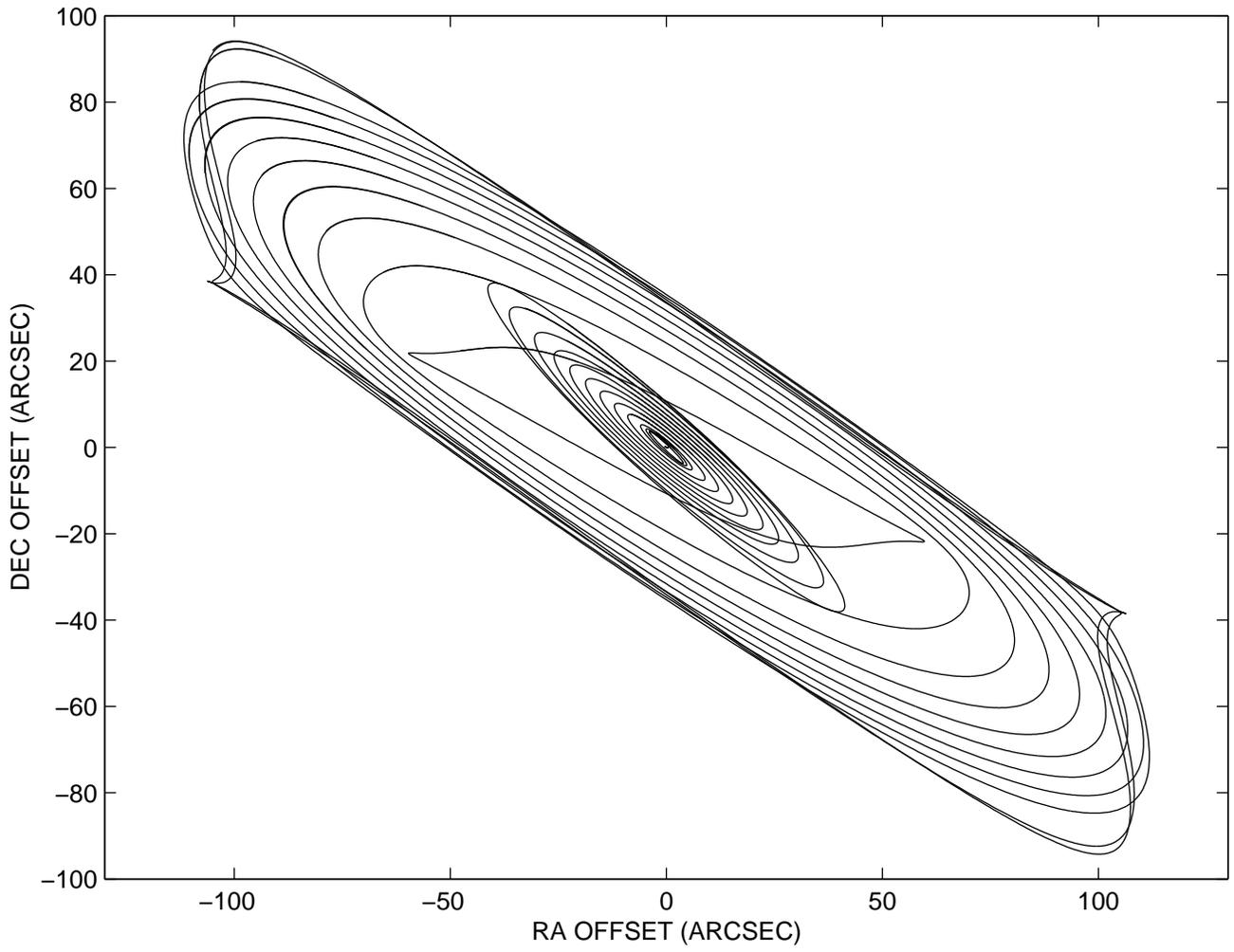
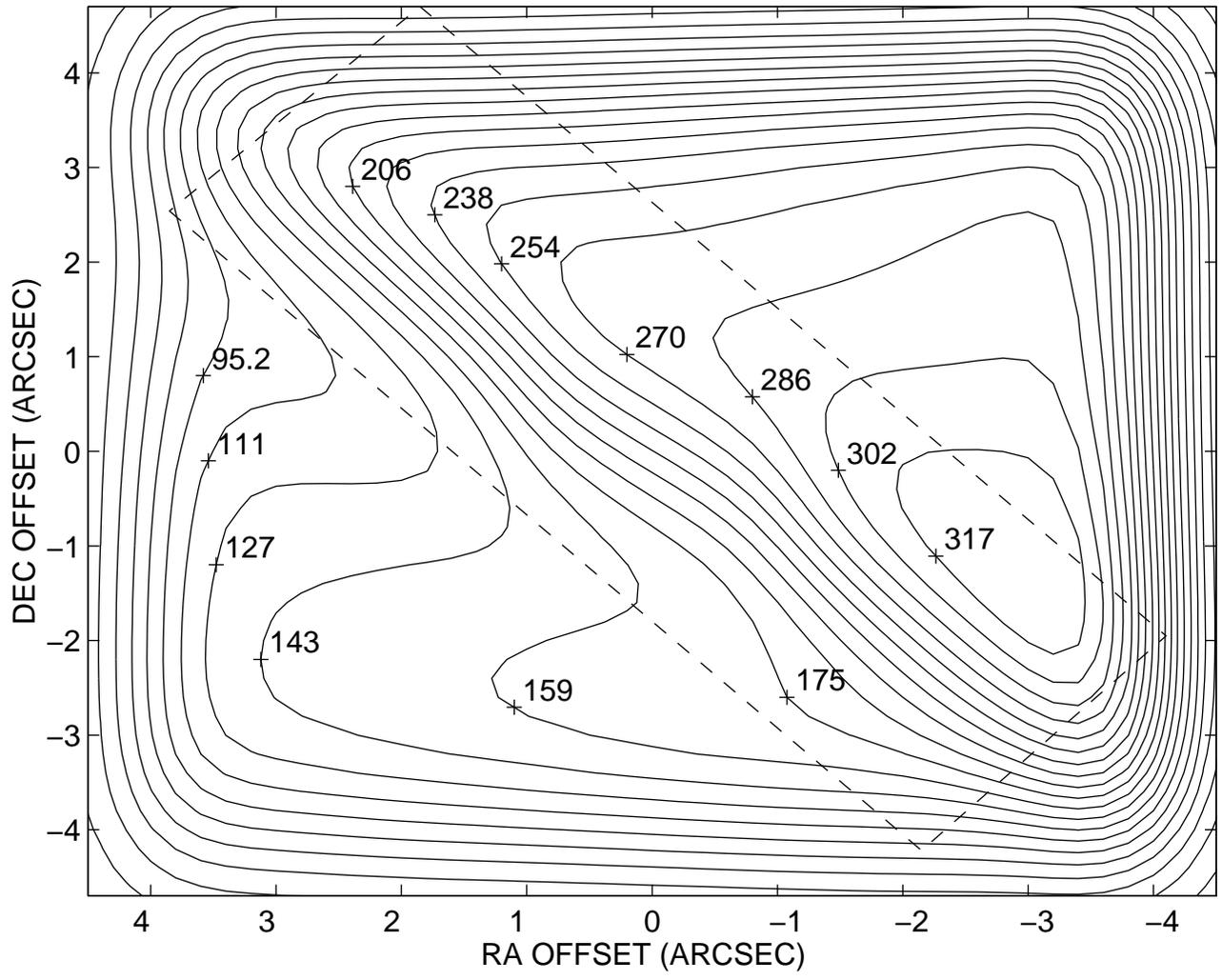
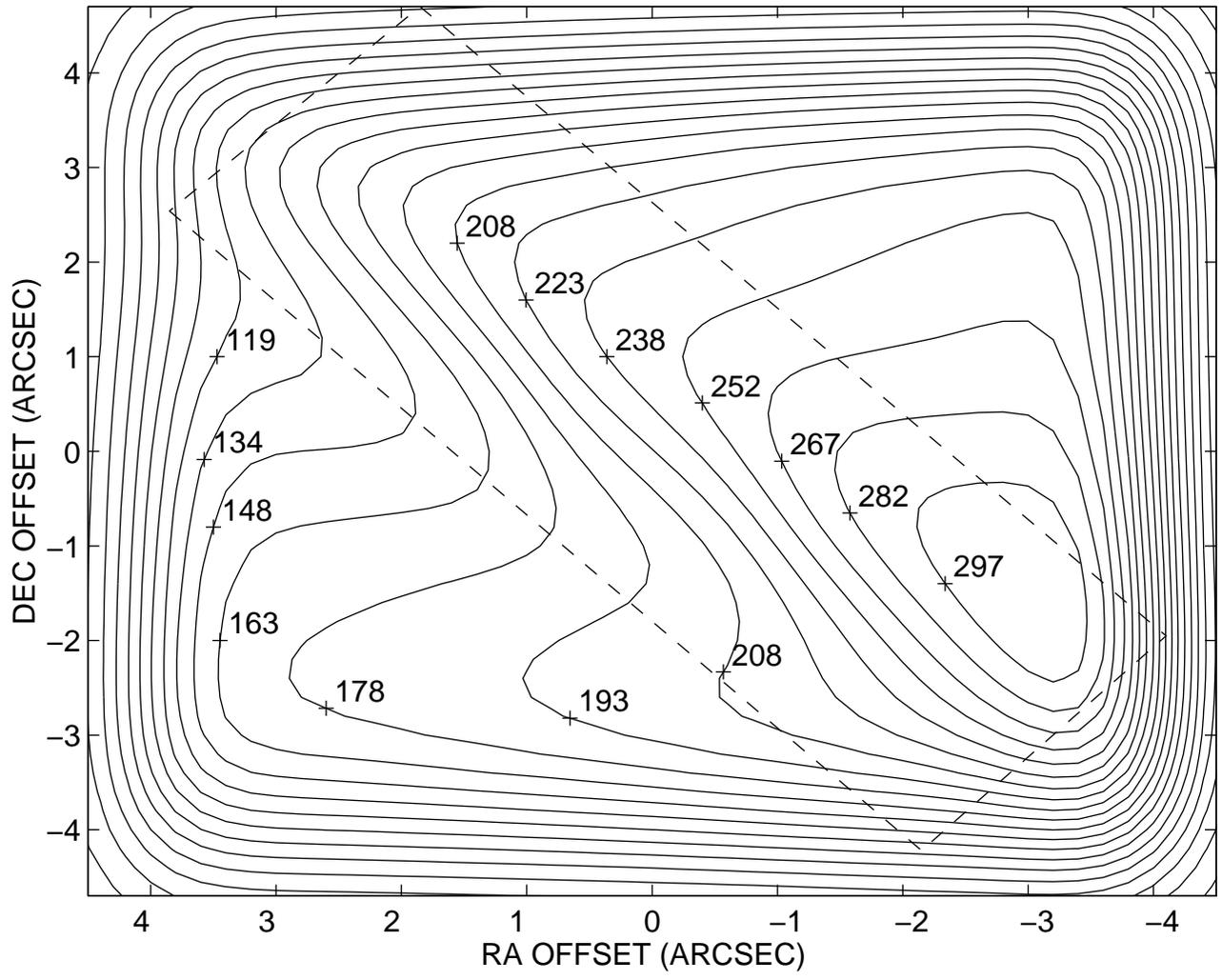
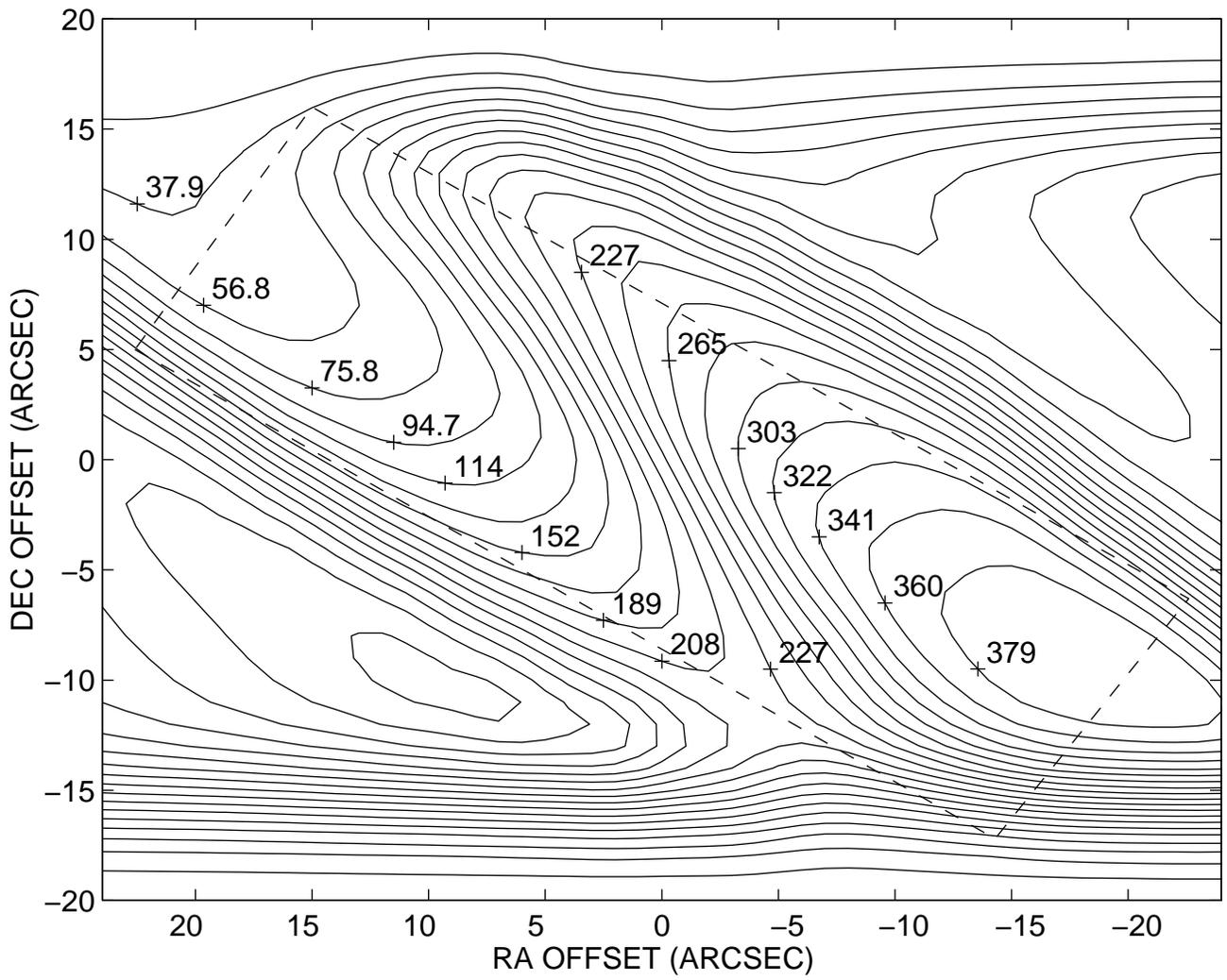
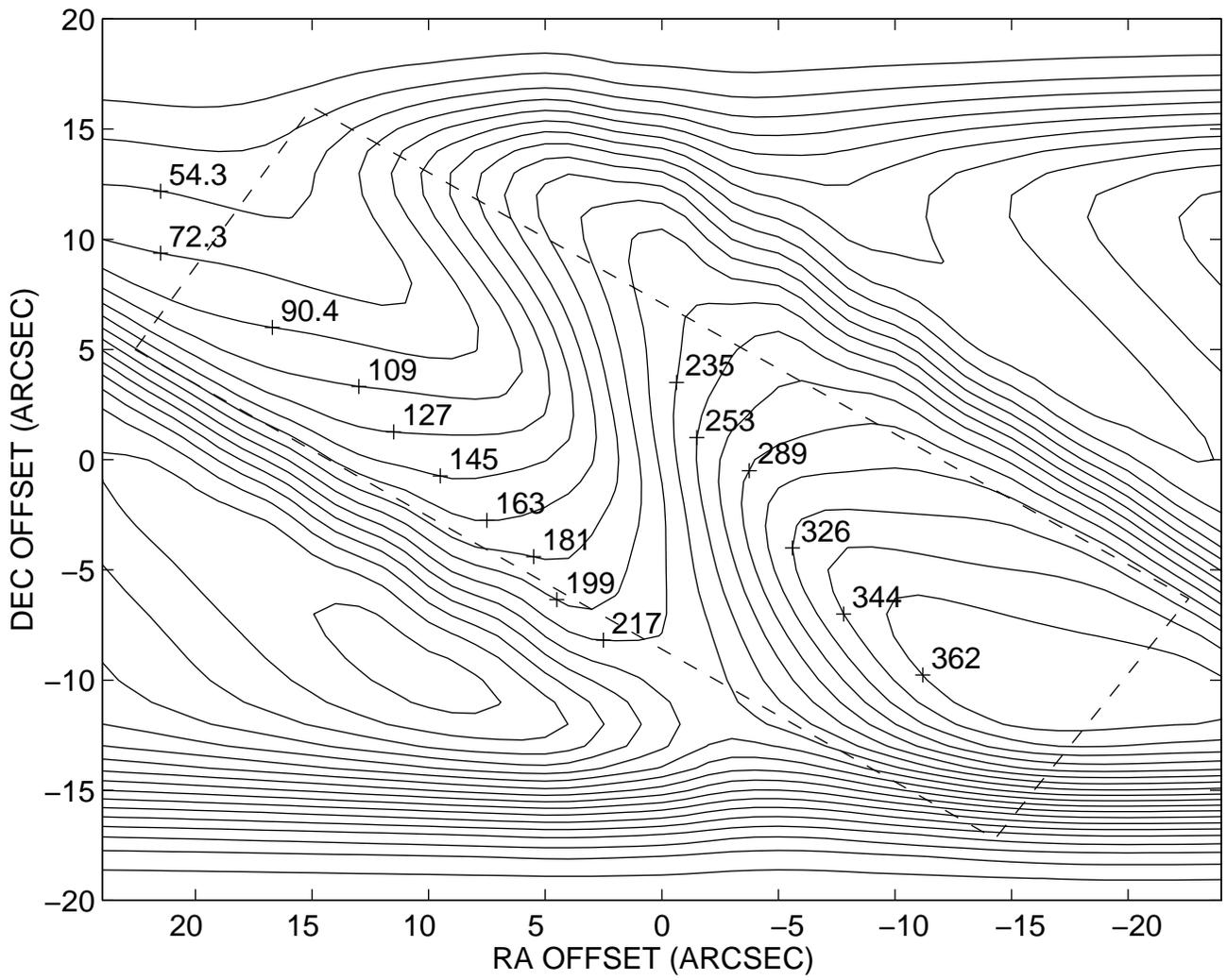
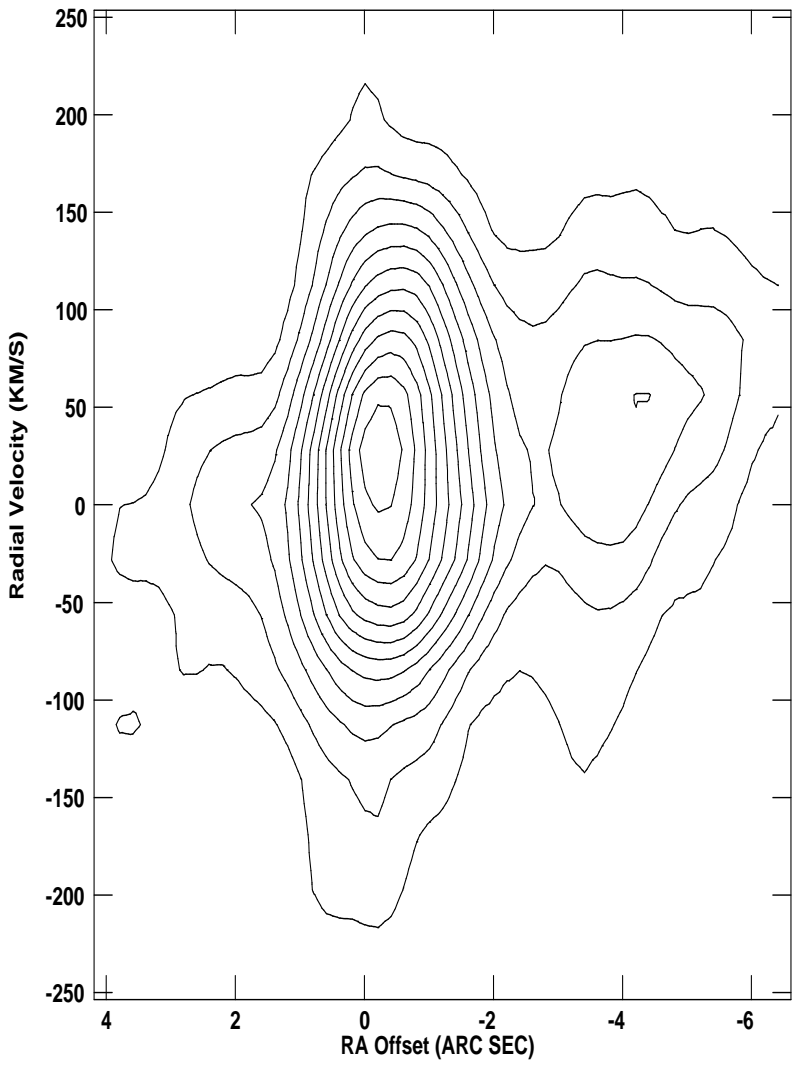
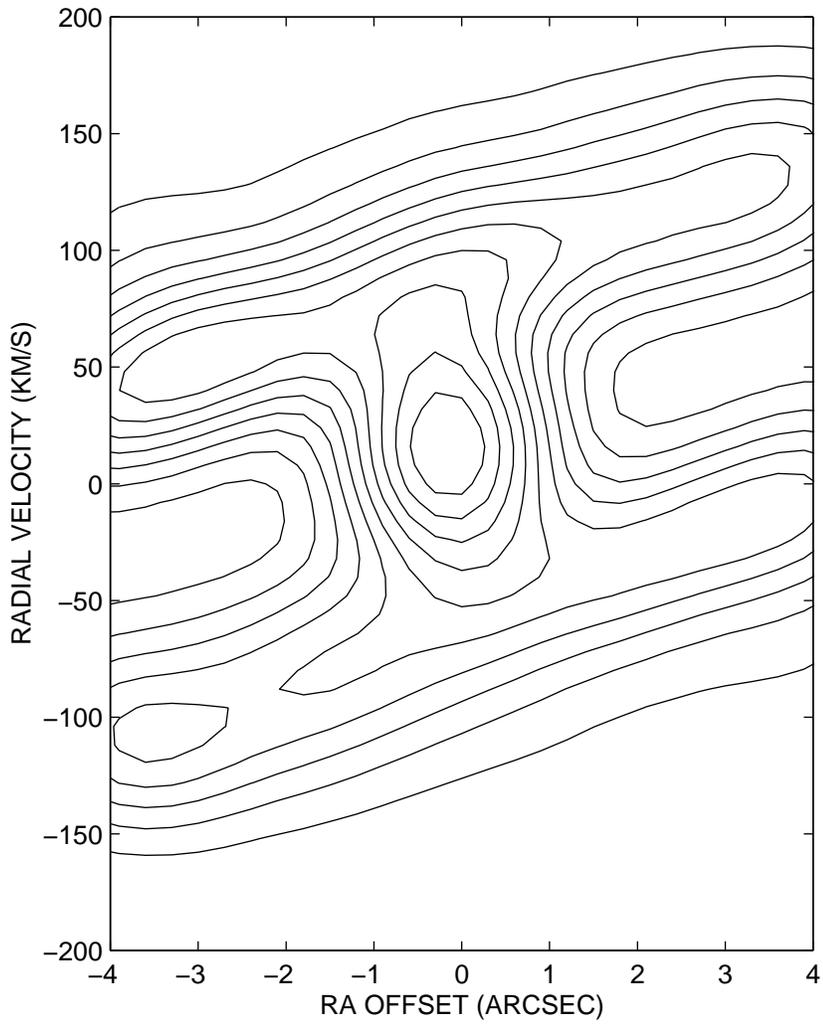
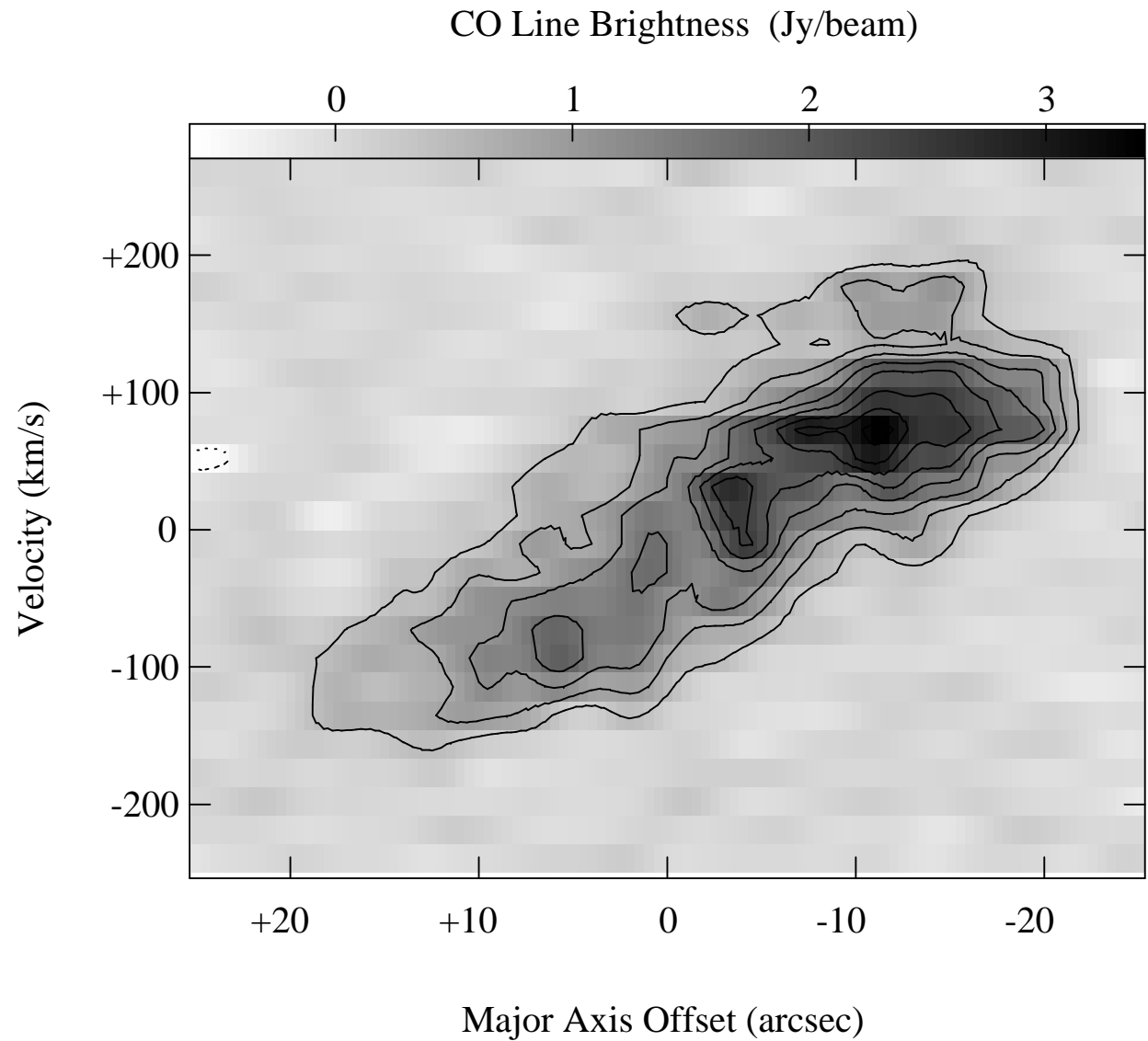
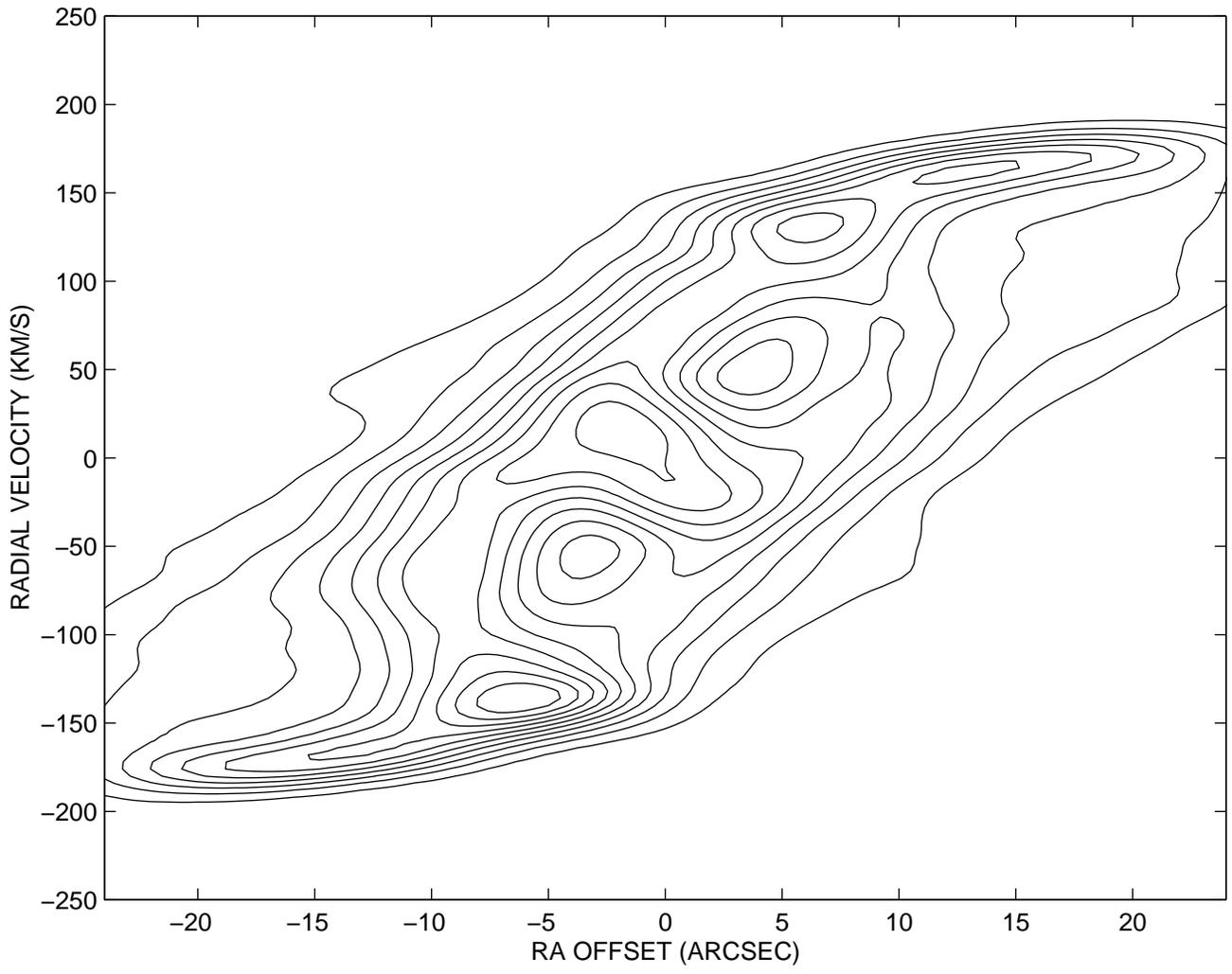
\end{document}